\documentclass[prl,twocolumn,showpacs,showkeys,amssymb]{revtex4}
\usepackage{graphicx}                

\begin{document}

\title{Strong clustering of non-interacting, passive sliders driven by a Kardar-Parisi-Zhang surface}

\author {Apoorva Nagar $^{1}$, Satya N. Majumdar $^{2}$ and Mustansir Barma $^{1}$}
\address{$^1$ Department of Theoretical Physics, Tata Institute of Fundamental Research, Homi Bhabha Road, Mumbai 400 005, India \\
$^2$ Laboratoire de Physique Theorique et Modeles Statistiques,\\ 
Universit'e Paris - Sud, B$\hat{a}$t 100, Orsay, France \\
}

\date{\today}

\begin{abstract}

We study the clustering of passive, non-interacting particles moving under the influence of a fluctuating field and random noise, in one dimension. The fluctuating field in our case is provided by a surface governed by the Kardar-Parisi-Zhang (KPZ) equation and the sliding particles follow the local surface slope. As the KPZ equation can be mapped to the noisy Burgers equation, the problem translates to that of passive scalars in a Burgers fluid. We study the case of particles moving in the same direction as the surface, equivalent to advection in fluid language. Monte-Carlo simulations on a discrete lattice model reveal extreme clustering of the passive particles. The resulting Strong Clustering State is defined using the scaling properties of the two point density-density correlation function. Our simulations show that the state is robust against changing the ratio of update speeds of the surface and particles. In the equilibrium limit of a stationary surface and finite noise, one obtains the Sinai model for random walkers on a random landscape. In this limit, we obtain analytic results which allow closed form expressions to be found for the quantities of interest. Surprisingly, these results for the equilibrium problem show good agreement with the results in the non-equilibrium regime. 

\end{abstract}

\pacs{05.40.-a,47.40.-x,02.50.-r,64.75.+g} 

\maketitle

\section{I. INTRODUCTION}

The coupling of two or more driven diffusive systems can give rise to intricate and interesting behavior, and this class of problems has attracted much recent attention. Models of diverse phenomena, such as growth of binary films ~\cite{drossel1}, motion of stuck and flowing grains in a sandpile ~\cite{biswas}, sedimentation of colloidal crystals  ~\cite{lahiri} and the flow of passive scalars like ink or dye in fluids ~\cite{shraiman,falkovich1} involve two interacting fields. In this paper, we concentrate on semiautonomously coupled systems --- these are systems in which one field evolves independently and drives the second field. Apart from being driven by the independent field, the passive field is also subject to noise, and the combination of driving and diffusion gives rise to interesting behavior. Our aim in this paper is to understand and characterize the steady state of a passive field of this kind.\\

The passive scalar problem is of considerable interest in the area of fluid mechanics and has been well studied, see ~\cite{shraiman,falkovich1} for reviews. Apart from numerical studies, considerable understanding has been gained by analyzing the Kraichnan model~\cite{kraichnan} where the velocity field of a fluid is replaced by a correlated Gaussian velocity field. Typical examples of passive scalars such as dye particles or a temperature field advected by a stirred fluid bring to mind pictures of spreading and mixing caused by the combined effect of fluid advection and diffusion. On the other hand, if the fluid is compressible, or if the inertia of the scalars cannot be neglected~\cite{falkovich}, the scalars may cluster rather than spread out. It has been argued that there is a phase transition as a function of the compressibility of the fluid --- at large compressibilities, the particle trajectories implode, while they explode in the incompressible or slightly compressible case~\cite{gawedzki}. It is the highly compressible case which is of interest in this paper.\\

Specifically, we study and characterize the steady state properties of passive, non-interacting particles sliding on a fluctuating surface and subject to noise ~\cite{drossel2, nagar}. The surface is the autonomously evolving field and the particles slide downwards along the local slope. We consider a surface evolving according to the Kardar-Parisi-Zhang (KPZ) equation. This equation can be mapped to the well known Burgers equation with noise, which describes a compressible fluid. Thus the problem of sliding passive particles on a fluctuating surface maps to the problem of passive scalars in a compressible fluid. We are interested in characterizing the steady state of this problem, first posed and studied by Drossel and Kardar in \cite{drossel2}. Using Monte-Carlo simulations of a solid on solid model and analyzing the number of particles in a given bin as a function of bin size, they showed that there is clustering of particles. However their analysis does not involve the scaling with system size, which as we will see below, is one of the most important  characteristics of the system. We find that the two point density-density correlation function is a scaling function of $r$ and $L$ ($r$ is the separation and $L$ is the system size) and that the scaling function diverges at small $r/L$. The divergence indicates formation of clusters while the scaling of $r$ with $L$ implies that the clusters are typically separated from each other by a distance that scales with the system size. A brief account of some of our our results has appeared in ~\cite{nagar}.\\ 

Scaling of the density-density correlation function with system size has also been observed in the related problem of particles with a hard core interaction, sliding under gravity on a KPZ surface ~\cite{das,das1,gopal1}. However, the correlation function in this case has a cusp singularity as $r/L \rightarrow 0$, in contrast to the divergence that we find for noninteracting particles. Thus, while clustering and strong fluctuations are seen in both, the nature of the steady states is different in the two cases. In our case, clustering causes a vanishing fraction of sites to be occupied in the noninteracting case, whereas hard core interactions force the occupancy of a finite fraction. In the latter case, there are analogies to customary phase ordered states, with the important difference that there are strong fluctuations in the thermodynamic limit, leading to the appellation Fluctuation Dominated Phase Ordering (FDPO) states. The terminology Strong Clustering States is reserved for the sorts of nonequilibrium states that are found with noninteracting particles --- a key feature being the divergent scaling function describing the two point correlation function.\\ 

\begin{figure}
  \centering
  \includegraphics[width=0.9\columnwidth,angle=0]{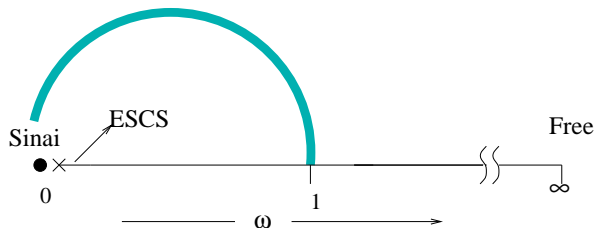}
  \caption{Change in state with change in $\omega$. In the $\omega \rightarrow 0$ limit one gets different kinds of states depending on how one approaches it. The $\omega \rightarrow \infty$ is the free particle limit. The arc shoes that there is a similarity between the results of the equilibrium Sinai limit and the non-equilibrium SCS at $\omega=1$.}
  \label{omega}
\end{figure}

In the problem defined above, there are two time scales involved, one associated with the surface evolution and the other with particle motion. We define $\omega$ as the ratio of the surface to the particle update rates. While we see interesting variations in the characteristics of the system under change of this parameter, the particular limit of $\omega \rightarrow 0$ is of special importance. There is a slight subtlety here as the limit $\omega \rightarrow 0$ does not commute with the thermodynamic limit $L \rightarrow \infty$. If we consider taking $\omega \rightarrow 0$ first and then approach large system size ($L \rightarrow \infty$), we obtain a state in which the surface is stationary and the particles move on it under the effect of noise. In this limit of a stationary surface, we obtain an equilibrium problem. This is the well known known Sinai model which describes random walkers in a random medium. We will discuss this limit further below. Now consider taking the large system size limit first and then approach $\omega = 0$; this describes a system in which particles move extremely fast compared to the evolution of the local landscape. This leads to the particles settling quickly into local valleys, and staying there till a new valley evolves. We thus see a non-equilibrium SCS state here, but with the features that the probability of finding a large cluster of particles on a single site is strongly enhanced. We call this limiting state the Extreme-strong clustering state (ESCS) (Fig.~\ref{omega}). The opposite limit shown in Fig.~\ref{omega} is the $\omega \rightarrow \infty$ limit where the surface moves much faster than the particles. Because of this very fast movement, the particles do not get time to ``feel'' the valleys and they behave as nearly free random walkers.\\

The equilibrium limit ($\omega \rightarrow 0$ followed by $L \rightarrow \infty$) coincides with the Sinai model describing random walkers in a random medium ~\cite{sinai}. This problem can be analyzed analytically by mapping it to a supersymmetric quantum mechanics problem  ~\cite{comtet} and we are able to obtain closed form answers for the two quantities of interest --- the two point correlation function $G(r,L)$ and the probability distribution function of finding $n$ particles on a site $P(n,L)$. Surprisingly, we find that not only do these results show similar scaling behavior as the numerical results for $\omega=1$ (nonequilibrium regime) but also the analytic scaling function describes the numerical data very well. The only free parameter in this equilibrium problem is the temperature and we choose it to fit our numerical data for the nonequilibrium system. Interestingly, the effective temperature seems to depend on the quantity under study.\\

The KPZ equation contains a quadratic term which breaks the up-down symmetry, thus one can have different behavior of the passive scalars depending on whether the surface is moving downwards (in the direction of the particles, corresponding to advection in fluid language) or upwards (against the particles, or anti-advection in fluid language). In this paper, we will consider only the case of advection. One can also consider dropping the nonlinear term itself; this leads to the Edwards-Wilkinson (EW) equation for surface growth. The problems of KPZ anti-advection and passive sliders on an Edwards Wilkinson surface are interesting in themselves and will be addressed in a subsequent paper ~\cite{future}.\\ 

Apart form the static quantities studied above, one can also study the dynamic properties of the system. Bohr and Pikovsky ~\cite{bohr} and Chin ~\cite{chin} have studied a similar model with the difference that they do not consider noise acting on particles. In the absence of noise, all the particles coalesce and ultimately form a single cluster in steady state, very different from the strongly fluctuating, distributed particle state under study here. References ~\cite{bohr} and ~\cite{chin} study the process of coalescence in time. Further, they find that the RMS displacement for a given particle increases in time $t$ as $t^{1/z}$, where $z$ is equal to the dynamic exponent of the surface, indicating that the particles have a tendency to follow the valleys of the surface. Drossel and Kardar ~\cite{drossel2} have studied the RMS displacement in the same problem in the presence of noise and observe the same behavior. We confirm this result in our simulations and observe that the variation of $\omega$ does not change the result.\\

The arrangement of this paper is as follows. In Section II, we will describe the problem in terms of continuum equations and then describe a discrete lattice model which mimics these equations at large length and time scales. We have used this model to study the problem via Monte Carlo simulations. Section III describes results of our numerical simulations. We start with results on the various static quantities in the steady state and define the SCS. We also report on the dynamic quantities and the effect on steady state properties of varying the parameter $\omega$. Section IV describes our analytic results for the equilibrium Sinai limit of a static surface and the surprising connection with results for the nonequilibrium problem of KPZ/Burgers advection.

\begin{figure}
  \centering
  \includegraphics[width=0.9\columnwidth,angle=0]{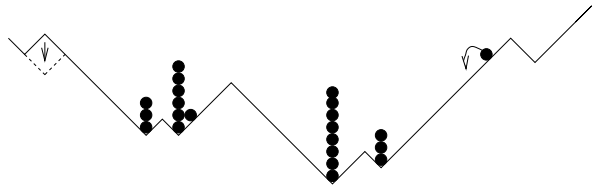}
  \caption{Schematic diagram of the surface and non-interacting particles sliding on top of it. Arrows show possible surface and particles moves.
}
  \label{picture1}
\end{figure}

\section{II. DESCRIPTION OF THE PROBLEM}
The evolution of the one-dimensional interface is described by the KPZ equation  \cite{kpz}  
\begin{eqnarray}
{\partial h \over \partial t} = \nu {\partial^{2} h \over \partial x^{2}}  + {\lambda \over 2} 
({\partial h \over\partial x})^2 + \zeta_h(x,t). 
\label{kpz1}
\end{eqnarray}
Here $h$ is the height field and $\zeta_h$ is a Gaussian white noise satisfying $\langle \zeta_h (x,t) \zeta_h({x}',t')\rangle = 2D_h \delta^d(x - {x}')\delta(t - t')$. For the passive scalars, if the $m^{th}$ particle is at position $x_m$, its motion is governed by  
 \begin{equation} 
  {dx_{m} \over dt} = \left.-a{\partial h \over dx } \right|_{x_{m}}+ \zeta_m (t)
  \label{passive} 
  \end{equation} 
where the white noise $\zeta_m (t)$ represents the randomizing effect
  of temperature, and satisfies $\langle \zeta_m (t) \zeta_m(t')\rangle
  = 2\kappa \delta(t - t')$. Equation~(\ref{passive}) is an overdamped Langevin equation of a particle in a potential $h(x,t)$ that is also fluctuating, with $a$ determining the speed of sliding. In the limit when $h(x,t)=h(x)$ is static, a set of noninteracting particles, at late times would reach the equilibrium Boltzmann state with particle density $\sim e^{-\beta h(x)}$. On the other hand, when $h(x,t)$ is time dependent, the system eventually settles into a strongly nonequilibrium steady state. The transformation $v = -\partial h / \partial x$ maps Eq.~(\ref{kpz1}) (with $\lambda = 1$) to the Burgers equation which describes a compressible fluid with local velocity $v$ 

 \begin{equation} 
  {\partial v \over \partial t} + \lambda(v.{\partial  \over \partial x})v=  \nu {\partial^2 v\over \partial x^2} + \ {\partial \zeta_h (x,t) \over \partial x}
  \label{burgers} 
  \end{equation} 
The above equation describes a compressible fluid because it does not have the pressure term, which is present in the Navier Stokes equation. The transformed Eq.~(\ref{passive}) describes passive scalar particles advected by the Burgers fluid 

 \begin{equation} 
  {dx_{m} \over dt} = \left.av \right|_{x_{m}}+ \zeta_m (t)
  \label{burgerspassive} 
  \end{equation} 
The ratio $a/ \lambda > 0$ corresponds to advection (particles moving with the flow), the case of interest in this paper, while $a/ \lambda < 0$ corresponds to anti-advection (particles moving against the flow).\\

Rather than analyzing the coupled Eqs.~(\ref{kpz1}) and ~(\ref{passive}) or equivalently Eqs.~(\ref{burgers}) and ~(\ref{burgerspassive}) directly, we study a lattice model  which is expected to have similar behavior at large length and time scales. The model consists of a flexible, one-dimensional lattice in which particles reside on sites, while the links or bonds between successive lattice sites are also dynamical variables which denote local slopes of the surface. The total number of sites is $L$. Each link takes either of the values $+1$ (upward slope $\rightarrow /$) or $-1$ (downward slope $\rightarrow \backslash$). The rules for surface evolution are : choose a site at random, and if it is on a local hill $(\rightarrow /\backslash)$, change the local hill to a local valley$(\rightarrow \backslash /)$ (Fig.~\ref{picture1}). After every $N_s$ surface moves, we perform $N_p$ particle updates according to the following rule : we choose a particle at random and move it one step downward with probability $(1+K)/2$ or upward with probability $(1-K)/2$.  The parameter $K$ ranges from 1 (particles totally following the surface slope) to 0 (particles moving independently of the surface). In our simulations, we update the surface and particles at independent sites, reflecting the independence of the noises $\zeta_h (x,t)$ and $\zeta_m (t)$ ~\cite{drosselcomment}. The ratio $\omega \equiv N_s/N_p$ controls the relative time scales of the surface evolution and particle movement. In particular, the limit $\omega \rightarrow 0$ corresponds to the adiabatic limit of the problem where particles move on a static surface and the steady state is the thermal equilibrium state.\\ 

To see how the lattice model described above describes a KPZ surface, consider the mapping of the above model to the well known asymmetric simple exclusion process (ASEP): consider an up slope to be a particle on a lattice and a down slope to be an empty space (hole). The flipping of a hill to a valley then corresponds to the motion of a particle (exchange of particle and hole). A coarse grained description of the ASEP leads to the KPZ equation ~\cite{barma}. The continuum description of the ASEP, obtained by coarse graining over regions which are large enough to contain many sites, involves the density of particles $\rho(x)$ and the local current $J(x)$. These are connected through the continuity equation 

\begin{eqnarray}
\frac{\partial \rho}{\partial t} + \frac{\partial J}{\partial x} = 0
\label{continuity}
\end{eqnarray}

The local current can be written as 

\begin{eqnarray}
J(x) = -\nu \frac{\partial \rho}{\partial x} + j(\rho) + \eta
\label{current}
\end{eqnarray}
where $\nu$ is the particle diffusion constant, $\eta$ is a Gaussian noise variable and $j(\rho)$ is the systematic contribution to the current associated with the local density $\rho$. Using the expression for the bulk ASEP with density $\rho$ for $j$, we have

\begin{eqnarray}
j(\rho)=(p-q)\rho(1-\rho)
\label{systematic}
\end{eqnarray}
where $p$ and $q$ are the particle hopping probabilities to the right and left respectively, with our one-step model corresponding to $p=1$ and $q=0$.\\

Since we identify the presence (absence) of a particle in the lattice model with an up (down) slope, we may write

\begin{eqnarray}
\rho=\frac{1}{2}(1+\frac{\partial h}{\partial x})
\label{connection}
\end{eqnarray}

Using Eqs.~(\ref{current}),(\ref{systematic}) and (\ref{connection}) in Eq.~(\ref{continuity}) leads to 

\begin{eqnarray}
{\partial h \over \partial t} = -\frac{1}{2}(p-q)+
\nu {\partial^{2} h \over \partial x^{2}} +  
\frac{1}{2}(p-q)({\partial h \over\partial x})^2 - \eta 
\label{kpz2}
\end{eqnarray}
which is the KPZ equation (Eq.~(\ref{kpz1})) with an additional constant term, and $\lambda=(p-q)$ and $\zeta_h=- \eta$. Note that the signs of the constant term and $\lambda$ are opposite. Thus a downward moving surface (corresponding to $p>q$) has positive $\lambda$. The constant term can be eliminated by the boost $h \rightarrow h-\frac{1}{2}(p-q)t$, but its sign is important in determining the overall direction of motion of the surface. The case $(a/\lambda) > 0$  which is of interest to us thus corresponds to the lattice model in which particles move in the same direction as the overall surface motion.\\     

The parameters $\omega$ and $K$ defined in the lattice model are connected to the continuum equations as follows. In the limit of a stationary surface, we achieve equilibrium and the particles settle into in a Boltzmann state with particle density $\sim e^{-\beta h(x)}$, here h(x) is the surface height profile and $\beta$ is the inverse temperature. $\beta$ is related to $K$ by $\beta=\ln\left(\frac{1+K}{1-K}\right)$ and to the parameters $a$ and $\kappa$ in Eq.~(\ref{passive}) by $\beta=a/\kappa$. Thus \begin{eqnarray}
K=\frac{e^{a/\kappa}-1}{e^{a/\kappa}+1}
\label{connect1}
\end{eqnarray}
 
 The parameter $\omega$ cannot be written simply in terms of the parameters in the continuum equations, because it modifies Eq.~(\ref{kpz1}) as we now show. $\omega$ is the ratio of the update speeds or equivalently the time between successive updates of the particles ($\Delta t_p$) and surface ($\Delta t_s$). The noises $ \zeta_h(x,t)$ and $\zeta_m(t)$ in Eqs. ~(\ref{kpz1}) and (\ref{passive}) can be written as $\sqrt{\frac{D_h}{\Delta t_s}}\widetilde{\zeta_h}(x,t)$ and $\sqrt{\frac{\kappa}{\Delta t_p}}\widetilde{\zeta_m}(t)$ respectively. Here $\widetilde{\zeta_h}(x,t)$ is noise of $O(1)$, uncorrelated in time, white in space while $\widetilde{\zeta_m}(t)$ is uncorrelated noise of $O(1)$. The factors of $\sqrt{\frac{1}{\Delta t}}$ in the terms indicate that the strength of the noise depends on how frequently noise impulses are given to the particles; the square root arises from the random nature of these impulses. Thus the change in height ($\Delta h$) in time $\Delta t_s$ and the distance traveled ($\Delta x_{m}$) in time $\Delta t_p$ are respectively -

\begin{eqnarray}
\Delta h = \Delta t_s[\nu {\partial^{2} h \over \partial x^{2}}  + {\lambda \over 2} ({\partial h \over\partial x})^2] + \sqrt{\Delta t_s D_h} \widetilde \zeta_h(x,t) 
\label{connect2}
\end{eqnarray}

\begin{equation} 
  \Delta x_m = \Delta t_p [\left.-a{\partial h \over \partial x_m } \right|_{x_m}]+ \sqrt{\Delta t_p \kappa} \widetilde \zeta(t)
\label{connect3} 
\end{equation} 
We now identify $\Delta t_s$ and $\Delta t_p$ with the Monte-Carlo time step $\delta t$ as $\Delta t_s=N_s \delta$ and $\Delta t_p=N_p \delta$. We can thus replace $\Delta t_s$ by $\omega . \Delta t_p$ and take it to be the natural continuous time. We thus get

\begin{eqnarray}
{\partial h \over \partial t} = \omega [\nu {\partial^{2} h \over \partial x^{2}}  + {\lambda \over 2} 
({\partial h \over\partial x})^2] + \sqrt{\omega} \zeta_h(x,t) 
\label{kpz3}
\end{eqnarray}

\begin{equation} 
  {dx_m \over dt} = \left.-a{\partial h \over dx_m } \right|_{x_m}+ \zeta_m(t)
\label{passive2} 
\end{equation}
We can see that the $\omega$ dependence in the above equation cannot be removed by a simple rescaling of the parameters of the equation. Eq.~(\ref{kpz1}) is recovered as a special case of Eq.~(\ref{kpz3}) on setting $\omega=1$.

\section{III. NUMERICAL RESULTS}

\subsection{Two Point Density Density Correlation Function}

\begin{figure}
  \centering
  \includegraphics[width=0.7\columnwidth,angle=-90]{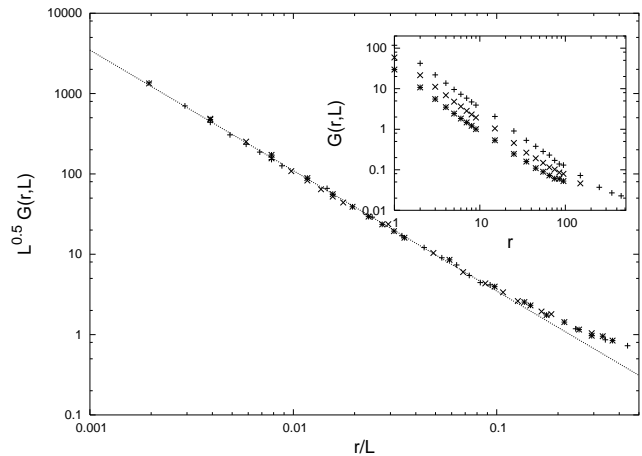}
  \caption{The inset shows $G(r,L)$ versus $r$ for different values of $L$. The main plot shows the scaling collapse when $r$ is scaled with $L$ and $G(r,L)$ with $1/L^{0.5}$. The dashed, straight line shows $y \sim x^{-1.5}$. The lattice sizes for both plots are $L$$=$  $256$ ($\ast$), $512$ ($\times$), $1024$ ($+$).
}
  \label{advncorr}
\end{figure}

We start with the simplest case $\omega = K = 1$; surface updates are attempted as frequently as particle updates, and both particles and surface always move only downwards. In our simulations, we work with $N=L$, where $N$ is the total number of particles and there are $L$ sites in the lattice. The two point density-density correlation function is defined as $G(r,L) = \langle n_i n_{i+r}\rangle_L$, where $n_i$ is the number of particles at site $i$. Fig.~\ref{advncorr} shows the scaling collapse of numerical data for various system sizes ($L$) which strongly suggests that for $r>0$, the scaling form
\begin{eqnarray} 
G(r,L) \sim \frac{1}{L^{\theta}} Y\left({\frac{r}{L}}\right) 
\label{correlation}
\end{eqnarray}    
is valid with $\theta \simeq {1/2}$. The scaling function $Y(y)$ has a power law divergence $Y(y) \sim y^{-\nu}$ as $y \rightarrow 0$, with $\nu$ close to 3/2. The data for $r=0$ points to $G(0,L) \sim L$.\\ 

This numerical result matches with an exact result of Derrida et. al. ~\cite{derrida} for a slightly different model. As we have seen in the previous section, the single step model which we use for Monte-Carlo simulations, can be mapped on to an asymmetric simple exclusion process (ASEP). The particles/holes in the ASEP map to the up/down slopes in our model and the flipping of a hill to a valley is equivalent to swapping a particle with a hole. In ~\cite{derrida}, apart from particles and holes, a third species called the second-class particles are introduced which act as holes for the particles and particles for the holes. When translated to the surface language, these second class particles behave like the sliders in our model, with the difference that they are not passive: there is no surface evolution at a site where second-class particles reside. The effect of non-passivity is relatively unimportant for KPZ advection-like dynamics of the surface, as particles mostly reside on stable local valleys while surface evolution occurs at local hilltops. Moreover, if the number of second class particles is small, the probability of the rare event where they affect the dynamics of local hills goes down even further. With only two such particles in the full lattice, probability $p(r)$ that they are at a distance $r$, is proportional to the two point correlation function $G(r,L)$. The exact result  ~\cite{derrida} $p(r) \sim \frac{1}{r^{3/2}}$ matches very well with our prediction for the same quantity, $p(r)=\frac{L}{N^{2}}G(r,L) \sim \frac{1}{r^{3/2}}$.\\

\begin{figure}
  \centering
  \includegraphics[width=0.7\columnwidth,angle=-90]{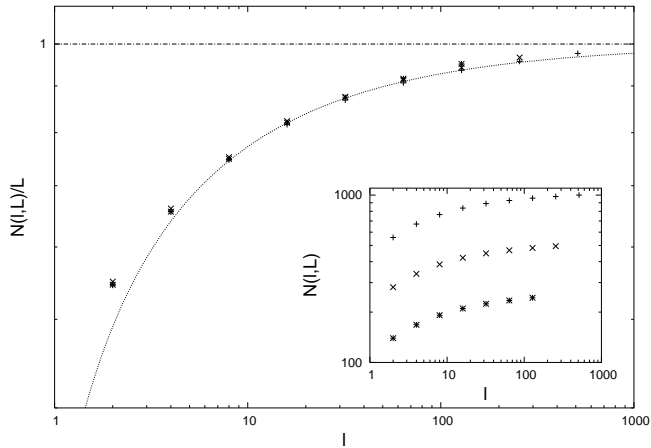}
  \caption{The inset shows  $N(l,L)$ versus bin size $l$ for different system sizes (L). The main plot shows $N(l,L)$ scaled with $L$ versus bin size $l$. The curve shows $c_{1}L(1-c_{2}l^{-\nu+1})$ with $c_{1}=1$ and $c_{2}=0.72$. The straight line shows $N(l,L)=L$, the form predicted in ~\cite{drossel2}. The lattice sizes for both plots are $L$$=$  $256$ ($\ast$), $512$ ($\times$), $1024$ ($+$).
}
  \label{advbin}
\end{figure}

The result for $G(r,L)$ also allows us to calculate the quantity $N(l,L)$ first defined in ~\cite{drossel2}; the lattice is divided into $L/l$ bins of size $l$ and we ask for the number $N(l,L)$ of particles in the same bin as a randomly chosen particle. $N(l,L)$ is a good measure of clustering - if $N(l,L)$ rises  linearly with $l$, one concludes that the particles are fairly spread out, while if $N(l,L)$ saturates or shows a decreasing slope, one concludes that particles are clustered. $N(l,L)$ is related to the two point correlation function through $N(l,L) = \int_0^l G(r,L) dr$, using which we obtain $N(l,L) \sim c_{1}L(1-c_{2}l^{-\nu+1})$. This form fits the numerical result for $N$ better (Fig.~\ref{advbin}) than the $l$-independent form of ~\cite{drossel2}.\\


\subsection{Probability Density of Occupancy}

\begin{figure}
  \centering
   \includegraphics[width=0.7\columnwidth,angle=-90]{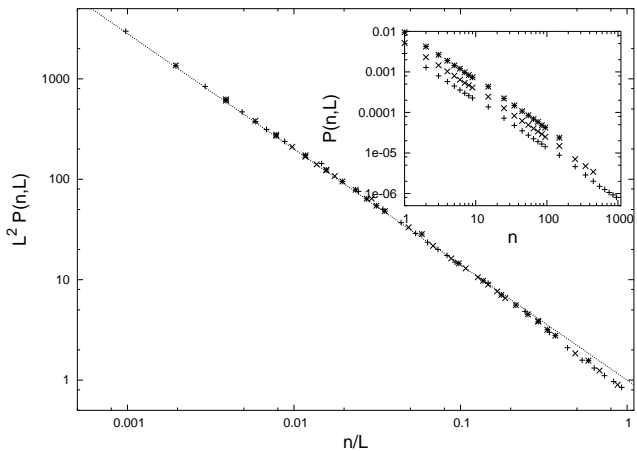}
  \caption{The inset shows  $P(n,L)$ versus $n$ for different values of $L$. The main plot shows $L^{2}P(n,L)$ versus $n/L$. The straight line shows $y \sim x^{-1.15}$. The lattice sizes are $L$$=$  $256$ ($\ast$), $512$ ($\times$), $1024$ ($+$).
}
  \label{advdensity}
\end{figure}

Another quantity of primary interest is the probability  $P(n,L)$ that a given site is occupied by $n$ particles. For $n>0$, this quantity shows a scaling with the total number of particles, which in turn is proportional to the system size $L$. We have (see Fig.~\ref{advdensity})
\begin{eqnarray}
P(n,L) \sim {1\over L^{2 \delta}} f \left({n\over L^{\delta}}\right), 
\label{probability}
\end{eqnarray}
with $\delta =1$. The scaling function $f(y)$ seems to fit well to a power law $y^{- \gamma}$ with $\gamma \simeq 1.15$ (Fig.~\ref{advdensity}), though as we shall see in Section IV, the small $y$ behavior may follow $y^{-1}lny$. We can use the scaling form in the above equation to calculate $G(0,L)$, $\langle n^2 \rangle \equiv G(0,L) = \int_0^Ln^{2}P(n,L)dn \sim L^{\delta} = L$, which, as we have seen above, is borne out independently by the numerics. Numerical data for $P(0,L)$ (which is not a part of the scaling function in  Eq.~(\ref{probability})) shows that  the number of occupied sites $N_{occ} \equiv (1-P(0,L))L$ varies as $L^{\phi}$ with $\phi \simeq 0.23$, though the effective exponent seems to decrease systematically with increasing system size $L$.\\ 

\begin{figure}
  \centering
   \includegraphics[width=0.7\columnwidth,angle=-90]{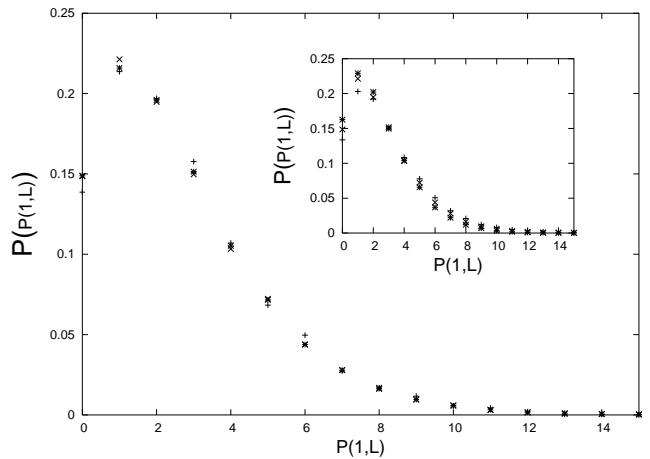}
  \caption{The inset shows  $\cal{P}$$(P(1,L))$ versus $P(1,L)$ for different values of L,  $L$$=$  $256$ ($\ast$), $512$ ($\times$), $1024$ ($+$). The main plot shows $\cal{P}$$(P(1,L))$ versus $P(1,L)$ for different number of averaging times,  $t$$=$  $T/10$ ($+$), $T$ ($\ast$), $T*5$ ($\times$) where T=30,000.
}
  \label{errdensity}
\end{figure}

\subsection{Fluctuations}

To evaluate the fluctuations of the quantity $P(n,L)$ about its mean, we evaluated the standard deviation --- $\Delta P(n,L) = \sqrt{\langle P(n,L)^2 \rangle - \langle P(n,L) \rangle^2}$ where the brackets denote an average over time. We find that this quantity does not decrease even in the thermodynamic limit $L \rightarrow \infty$. Let us ask for the probability density function $\cal{P}$$(P(n,L))$ describing the values taken on by $P(n,L)$. As seen in (Fig.~\ref{errdensity}), this distribution does not change when we increase the averaging time (main figure) or the length (inset). Thus  $\cal{P}$$(P(n,L))$ approaches a distribution with a finite width in the thermodynamic limit rather than a delta function. This clearly indicates that there are large fluctuations in the system which do not damp out in the thermodynamic limit. Large fluctuations, which do not decrease with increasing system size, are also a feature of the FDPO state for particles with a hard core interaction ~\cite{das,das1,gopal1}.

\subsection{Results on Dynamics}

The root mean square (RMS) displacement $R(t)=\langle (x(t)-x(0))^{2} \rangle^{1/2}$ of a tagged particle has been studied earlier ~\cite{chin, drossel2}. $R(t)$ is found to obey the scaling form

\begin{eqnarray}
R(t)= L^{\chi}h\left({t\over L^{z}}\right)
\label{rms1}
\end{eqnarray}
where $h(y) \sim y^{1/z}$, with $z=3/2$ for small $y$. The requirement that $R(t)$ has to be independent of $L$ in the limit $L \rightarrow \infty$ leads to $\chi=1$. The value of $z$ above is the same as the dynamic exponent of the KPZ surface. The dynamic exponent $z_s$ of a surface carries information about the time scale of evolution of valleys and hills; the landscape evolves under surface evolution and valleys/hills of breadth $L'$ are typically replaced by  hills/valleys in time of order $L'^{z_s}$. Thus the observation $z=z_s$ suggests that the particles follow the valley movement.\\ 

We have also evaluated the autocorrelation function $\widetilde{G}(t,L) \equiv \langle n_i(0) n_{i}(t)\rangle_{L}$ and find that it scales with the system size as 

\begin{eqnarray} 
 \widetilde{G}(t,L) \sim \widetilde{Y} \left(t \over L^{z}\right).
\label{autocorrelation}
\end{eqnarray}
Again, $z = z_s = 3/2$, reaffirming our conclusion that particles tend to follow valleys. The scaling function shows a power law behavior $ \widetilde{Y}(\tilde y)\sim \tilde y^{- \psi}$ with $\psi \simeq 2/3$ as $\tilde y \rightarrow 0$.\\ 

\subsection{Relations Between the Exponents}

The exponents defined in the above sections can be connected to each other by simple relations using scaling analysis. For instance, $\delta$, $\nu$ and $\theta$ are related by

\begin{eqnarray}
 \delta = \nu - \theta
\label{exponent1}
\end{eqnarray}
This can be proved by substituting the scaling form of Eq.~(\ref{correlation}) and $G(0,L) = \int_0^Ln^{2}P(n,L)dn \sim L^{\delta}$ in the equation $\int_0^LG(r,L)dr = L$; the last equation can be obtained by using the definition of $G(r,L)$ and using $N=L$. We can also relate $\phi$, $\delta$ and $\gamma$ by 

\begin{eqnarray}
 \phi = \delta(\gamma-2)+1
\label{exponent2}
\end{eqnarray}          
which can be derived using the normalization condition $\int_0^LP(n,L)dr = 1$ and then substituting for $P(0,L)$ and the scaling form of Eq.~(\ref{probability}). Our results from simulations are consistent with these relations.\\

The following picture of the steady state emerges from our results. The scaling of the probability distribution $P(n,L)$ as $n/L$ and the vanishing of the probability of finding an occupied site ($\equiv N_{occ}/L$) suggest that a large number of particles (often of the order of system size) aggregate on a few sites. The scaling of the two-point density-density correlation function with $L$ implies that the particles are distributed over distances of the order of $L$, while the divergence of the scaling function indicates clustering of large-mass aggregates. Thus the evidence points to a state where the particles form a few, dense clusters composed of a small number of large mas aggregates and these clusters are separated on the scale of system size. We choose to call this state as the Strong Clustering State (SCS). The divergence at origin of the two-point density-density correlation function as function of the separation scaled by the system size, is its hallmark. The information we get from results on dynamics is that the particles have a tendency to follow the surface. This is brought out by the fact that the scaling exponent describing the RMS displacement comes out to be equal to the dynamic exponent of the KPZ surface.\\ 

\subsection{Variation of $\omega$ and $K$}

\begin{figure}
  \centering
  \includegraphics[width=0.7\columnwidth,angle=-90]{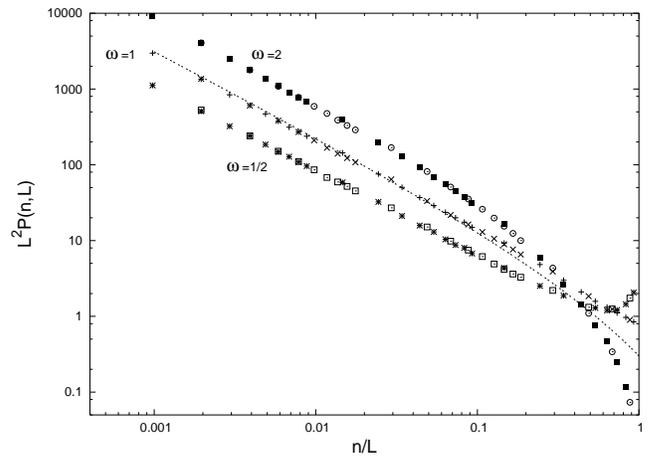}
  \caption{Scaled probability distribution $P(n,L)$ for $\omega = 1/2,1,2$ $(K=1)$. The line is a fit to Eq.~(\ref{gy1}) with $\beta=2.3$. The lattice sizes are $L$$=$ $512$ ($\circ$, $\times$, $\Box$), $1024$ ($\blacksquare$, $+$, $\ast$).}
  \label{advdensityomega}
\end{figure}

\begin{figure}
  \centering
  \includegraphics[width=0.7\columnwidth,angle=-90]{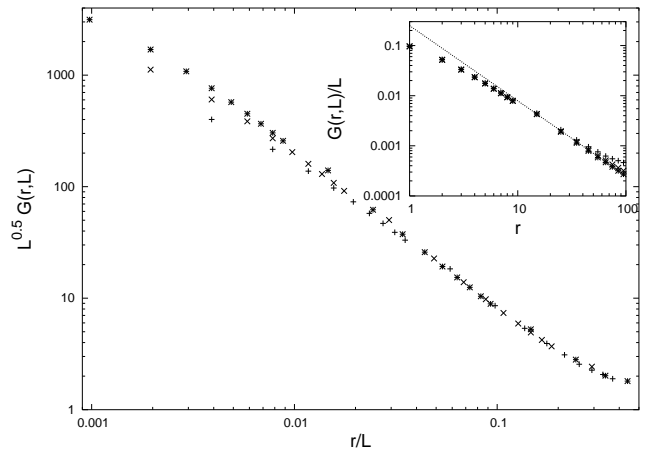}
  \caption{The main plot shows the scaled two point correlation function for $\omega = 2$, $(K=1)$, we see deviation from scaling at small $r/L$. The inset shows a plot of $G(r,L)/L$ versus $r$. The straight line shows depicts the power law $y \sim x^{-1.5}$. The lattice sizes are $L$$=$ $256$ ($+$),$512$ ($\times$), $1024$ ($\ast$).}
  \label{advcorrslow}
\end{figure}

To see how the system behaves when we change the relative speeds of the surface and particle evolution, we vary the parameter $\omega \equiv N_s/N_p$ ($N_s$ and $N_p$ being respectively the number of successive surface and particle update attempts) in the range $1/4 \leq \omega \leq 4$. When $\omega < 1$ (particles faster than the surface), we regain the scaling form of Eq.~(\ref{correlation}) for the two point correlation function. The scaling function also diverges with the same exponent. While the probability distribution for occupancy $P(n,L)$ shows similar scaling with system size as Eq.~(\ref{probability}), the scaling function $f(y)$ shows a new feature --- it develops a peak at large $n$ (Fig.~\ref{advdensityomega}). This peak at large $n$ indicates that the probability of finding nearly all the particles at a single site is substantial. A heuristic argument for the appearance of this peak is now given. Consider a configuration in which a large number of particles (nearly equal to the total number of particles) reside in a local valley. When this valley is replaced by another one nearby under surface dynamics , all the particles tend to move to the new one. If the number of particle updates is greater than surface updates, there is a substantial probability that all the particles are able to move to the new valley before it is itself replaced by another one. Thus there is a significant probability of the initial cluster surviving intact for a fair amount of time. Numerically, we also find that 

\begin{eqnarray} 
 \frac{P(n=N)}{P(n=N-1)}=\frac{1}{\omega}
\label{advprobomega}
\end{eqnarray}

For $\omega > 1$, the particles settle down slowly in valleys and $\tau_{surf} \gg \tau_{part}$ where $\tau_{surf}$ and $\tau_{part}$ are respectively the times between successive surface and particle updates. Though $\tau_{surf} \gg \tau_{part}$; for large enough $L$, the survival time of the largest valley $\sim \tau_{surf} L^z$ is always greater than the particle sliding time  $\sim \tau_{part} L$. Thus we expect that particles will lag behind the freshly formed valleys of small sizes but would manage to cluster in the larger and deeper valleys, which survive long enough. We thus expect a clustering of particles and scaling to hold beyond a crossover length scale ($r_c(\omega)$). We can estimate the crossover length by equating the time scales of surface and particle rearrangements --- $\tau_{surf} r_c^z(\omega) \sim \tau_{part} r_c(\omega)$, which yields $r_c(\omega)\sim \omega^{\frac{1}{z-1}}$. Using $z=3/2$, we have $r_c \sim \omega^2$. Numerical simulation results are shown in Fig.~\ref{advcorrslow} which shows that the data deviates from scaling and power law behavior at small $r$, due to the crossover effect.
The data suggests that 

\begin{eqnarray} 
G(r,L) \sim \sim \frac{1}{L^{\theta}} Y\left({\frac{r}{L}}\right)g(\frac{r}{r_c(\omega)}) 
\label{advcorrelationslow}
\end{eqnarray}   
As we can see from Fig.~\ref{advcorrslow} (main graph), the curve flattens out at small values of $r$, so for $y<1$ ($r<r_c(\omega)$), the function $g(Y)$ in the equation above should follow $g(y) \sim y^{1.5}$ while it should go to a constant for $y>1$. We can determine $r_c(\omega)$ from $G(r,L)$ by separating out the $r$ dependent part; if we scale $G(r,L)$ by $L$, we obtain the quantity $\frac{1}{r^{1.5}}g(\frac{r}{r_c(\omega}))$. We can now determine $r_c(\omega)$ as the value or $r$ where the scaled data starts deviating from the power law behavior $r^{-1.5}$. From Fig.~\ref{advcorrslow}, (inset) $r_c(\omega=2) \simeq 10$. A similar exercise for $\omega=3$ leads to $r_c(\omega=3) \simeq 20$. A clean determination of $r_c(\omega)$ for $\omega>3$ requires data for very large values of system size, beyond the scope of our computational capabilities.\\ 

The probability distribution $P(n,L)$ continues to show the same scaling form (Eq.~(\ref{probability})) for $\omega>1$, but the scaling function $f(y)$ in this case dips at large values of $y$ (Fig.~\ref{advdensityomega}) in contrast to the peak seen for $\omega<1$. The exponent $z$ describing the RMS displacement of particles remains unchanged under a change in $\omega$, again indicating that particles follow the movement of valleys on the large scale.\\

\begin{figure}
  \centering
  \includegraphics[width=0.7\columnwidth,angle=-90]{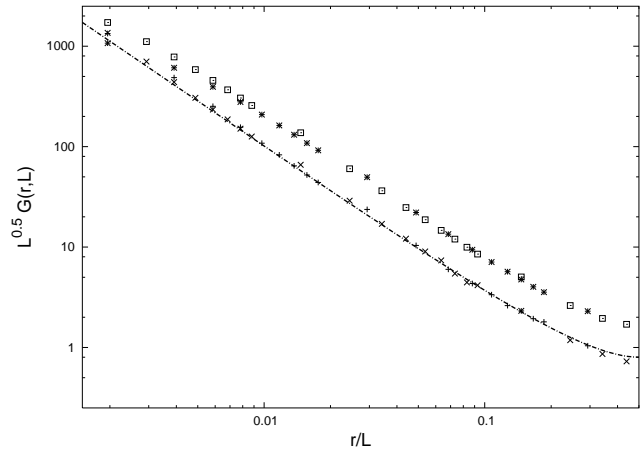}
  \caption{Two point scaled density correlation $G(r,L)$ function (advection) for $\omega=1$, $K=0.75$ (top curve), $1$ (bottom). The line is a plot of Eq.~(\ref{connection2}) with $\beta = 4$. The lattice sizes are $L~=~~1024$ ($\Box$, $\times$), $512$ ($\ast$, $+$).
}
  \label{advcorr}

\end{figure}

The other parameter of interest is $K$, defined in Section II --- when we make a particle update, we move the particle downhill with probability $(1+K)/2$ and uphill with probability $(1-K)/2$. So far we have discussed the results for the case $K=1$, where particles always move downhill. Decrease in $K$ reduces the average downhill speed of particles, while the valley evolution rates are unaffected. Thus decreasing $K$ causes an effect similar to increasing $\omega$ and a crossover length scale is introduced. The particles lag behind the freshly formed local valleys but settle down in the deeper, longer surviving global valleys. The numerical results again guide us to the form  
\begin{eqnarray} 
G(r,L) \sim \frac{1}{L^{\theta}} Y\left({\frac{r}{L}}\right)g(\frac{r}{r_c(K)}) 
\label{advcorrelationslow1}
\end{eqnarray} 
for the correlation function. Analogous to $\omega > 1$ case, we have extracted $r_c$ from the numerical data. We find $r_c(K=0.75) \simeq 10$. Values of $K$ lower than $0.75$ require data for system sizes that are beyond our computational limitations.\\

\section{IV. ADIABATIC, EQUILIBRIUM LIMIT}

To approach the problem analytically we take the extreme limit of $\omega \rightarrow 0$. In this limit, the surface is stationary. The particles move on this static surface under the effect of noise and the problem becomes a well known equilibrium problem - the Sinai model ~\cite{sinai} for random walkers on a random landscape. It is well known that for the KPZ surface in one-dimension, the distribution of heights $h(r)$ in the stationary state is described by 

\begin{eqnarray} 
{\rm Prob} [\{h(r)\}]\propto \exp\left[-{\nu \over {2D_h}}\int \left(\frac{dh(r')}{dr'}\right)^2 dr'\right]
\label{equisurface}
\end{eqnarray}   
Thus, any stationary configuration can be thought of as the trace of a random walker in space evolving via the equation, $dh(r)/dr = \xi(r)$ where the white noise $\xi(r)$ has zero mean and is delta correlated, $\langle \xi(r)\xi(r')\rangle = \delta (r-r')$. We impose periodic boundary conditions as for the lattice model, without loss of generality - $h(0)=h(L)=0$.\\ 

The passive particles moving on top of this surface as, we remember, move according to  Eq.~(\ref{passive}). Since this is an equilibrium situation,  $\langle \zeta_m (t) \zeta_m(t')\rangle = 2\kappa \delta(t - t') = 2K_{B}T \delta(t - t')$ where $T$ is the temperature and $K_{B}$ is the Boltzmann constant. Since  the particles are non-interacting, we can deal with a single particle and instead of the number of particles $n_{r}$ at a site $r$, we consider the probability $\rho(r)dr$ that the particle will be located between $r$ and $r+dr$. In the long time time, the particle will reach its thermal equilibrium in the potential $h(r)$ and will be distributed according to the Gibbs-Boltzmann distribution,
\begin{equation}
\rho(r) = { {e^{-\beta h(r)}}\over {Z}},
\label{Gibbs1}
\end{equation}
where $Z=\int_0^L dr e^{-\beta h(r)}$ is the partition function. Note that $\rho(r)$ in Eq. (\ref{Gibbs1}) depends on the realization of the potential $\{h(r)\}$ and varies from one realization to another. Our goal would be to compute the distribution of $\rho(r)$ over different realization of the random potential $h(r)$ drawn from the distribution in Eq. (\ref{equisurface}). Note that the distribution of $h(r)$ in Eq. (\ref{equisurface}) is invariant under the transformation $h(r)\to -h(r)$. In other words, the equilibrium density $\rho(r)$ defined in Eq. (\ref{Gibbs1}) will have the same distribution if one changes the sign of $h(r)$ in Eq. (\ref{Gibbs1}). For later simplicity, we will make this transformation now and replace $\rho(r)$ instead by the following definition
\begin{equation}
\rho(r) =  { {e^{\beta h(r)}}\over {Z}},
\label{Gibbs2}
\end{equation}
where the transformed partition function is now given by $Z=\int_0^L dr e^{\beta h(r)}$.

\subsection{ The Exact Distribution of the Probability Density}

Our strategy would be first to compute the $n$-th moment of the random variable $\rho(r)$ in Eq. (\ref{Gibbs2}). It follows from Eq. (\ref{Gibbs2}) - 

\begin{equation}
{\rho^n(r)} = { {e^{n\beta h(r)}}\over {Z^n}}={1\over {\Gamma (n)}}\int_0^{\infty} dy\, y^{n-1}e^{-yZ + n \beta h(r)}
\label{nmom1}
\end{equation}

where we have used the identity $\int_0^{\infty} dy\, y^{n-1} e^{-yZ} = \Gamma(n)/Z^n$ to rewrite the factor $1/Z^n$. Here $\Gamma(n)$ is the standard Gamma function. Next, we make a further change of variable in Eq. (\ref{nmom1}) by writing $y = \beta^2 e^{\beta u}/2$. Note that as $y$ varies from $0$ to
$\infty$, the new dummy variable $u$ varies from $-\infty$ to $\infty$. Making this substitution in
Eq. (\ref{nmom1}) we get,
\begin{eqnarray}
{\rho^n(r)} & = & b_n \int_{-\infty}^{\infty} du\, \exp [-{{\beta^2}\over {2}}\left\{\int_0^{L} dx\, e^{\beta(h(x)+u)} \right \} \nonumber\\ 
 & & + n\beta (h(r)+u) ] 
\label{nmom2}
\end{eqnarray}
where we have used the explicit expression of the partition function, $Z=\int_0^L dr e^{\beta h(r)}$. The constant $b_n = \beta^{2n+1}/[2^n \Gamma(n)]$. We are now ready to average the expression in Eq. (\ref{nmom2}) over the disorder, i.e., all possible realizations of the random potential $h(x)$ drawn from the distribution in Eq. (\ref{equisurface}). Taking the average in Eq. (\ref{nmom2}) (we 
denote it by an overbar), we get using Eq. (\ref{equisurface}),
\begin{eqnarray}
{\overline {\rho^n(r)}} & = &A b_n \int_{-\infty}^{\infty} du \int_{h(0)=0}^{h(L)=0} {\cal D} h(x) \exp [-\left\{  \int_0^{L} dx \right.  \nonumber \\  
 & & \left.  
\left[{1\over {2}}{\left( {{dh(x)}\over
 {dx}}\right)}^2 +{{\beta^2}\over {2}}e^{\beta(h(x)+u)}\right]\right\} \nonumber \\ 
 & & + n\beta(h(r)+u)]
\label{avmom1}
\end{eqnarray}
where the normalization constant $A$ of the path integral in Eq. (\ref{avmom1}) will be chosen so as to satisfy the sum rule, $\int_0^{L}{\overline {\rho(r)}}dr=1$. Next we shift the potential by a constant amount $u$, i.e., we define a
new function $V(x)= h(x)+u$ for all $x$ that reduces Eq. (\ref{avmom1}) to the following expression,
\begin{eqnarray}
{\overline {\rho^n(r)}} & = &A\, b_n \int_{-\infty}^{\infty} du\,\int_{V(0)=u}^{V(L)=u} {\cal D} V(x)\,
\exp\left[-\left\{\int_0^{L} dx\, \right. \right. \nonumber \\ 
 & & \left. \left. \left[{1\over {2}}{\left( {{dV(x)}\over
{dx}}\right)}^2 +{{\beta^2}\over {2}}e^{\beta V(x)}\right]\right\}
+ n\beta V(r) \right]
\label{avmom2}
\end{eqnarray} 
This path integral can be viewed as a quantum mechanical problem in the following sense. All paths (with the measure shown above) starts from $V(0)=u$ and ends at $V(L)=u$. At the fixed point $r$ (where we are trying to calculate the density distribution), these paths take a value $V(r)=V$ which can vary from $-\infty$ to $\infty$. This can be written, using the quantum mechanical bra-ket notation,
\begin{eqnarray}
{\overline {\rho^n(r)}} & = &A\, b_n \int_{-\infty}^{\infty} du\, \int_{-\infty}^{\infty} dV <u|e^{-{\hat H} r}|V> e^{n\beta V} \nonumber \\
 & & <V| e^{-{\hat H}(L-r)}|u>
\label{pi1}
\end{eqnarray}
The first bra-ket inside the integral in Eq. (\ref{pi1}) denotes the propagation of paths from the initial value $u$ to $V$ at the intermediate point $r$ and the second bra-ket denotes the subsequent propagation of the paths from $V$ at $r$ to the final value $u$ at $L$. The Hamiltonian $\hat H$ corresponds to the operator ${\hat H}\equiv {1\over {2}}\left({{dV}\over {dx}}\right)^2 + {{\beta^2}\over {2}}e^{\beta V(x)}$. Interpreting $V(x)$ to be the ``position" of a fictitious particle at the fictitious ``time" $x$, this operator has a a standard kinetic energy term and a potential energy which is exponential in the ``position" $V$. The right hand side of Eq. (\ref{pi1}) can be rearranged and simplified as in the following -
\begin{eqnarray}
{\overline {\rho^n(r)}} & = &A\, b_n \int_{-\infty}^{\infty} dV\, e^{n \beta V} \int_{-\infty}^{\infty} du <V|e^{-{\hat H}(L-r)}|u> \nonumber \\
 & & <u|e^{-{\hat H} r}|V>
\label{pi2}
\end{eqnarray}
Thus,
\begin{eqnarray}
{\overline {\rho^n(r)}}&=& A\, b_n \int_{-\infty}^{\infty} dV\, e^{n \beta V} <V|e^{-{\hat H} L}|V>
\label{pi2a}
\end{eqnarray}
where we have used the completeness condition, $\int_{-\infty}^{\infty} du\, |u><u| = {\hat I}$ with $\hat I$ being the identity operator.At this point, it may be helpful and less confusing notationally if we denote the ``position" $V$
of the fictitious quantum particle by a more friendly notation $V\equiv X$, which will help us thinking more clearly. Thus, the Eq. (\ref{pi2a}) then reduces to,\begin{equation}
{\overline {\rho^n(r)}}
= A\, b_n \int_{-\infty}^{\infty} dX\, e^{n \beta X} <X|e^{-{\hat H} L}|X>.
\label{pi3}
\end{equation}
To evaluate the matrix element in Eq. (\ref{pi3}), we need to know the eigenstates and the eigenvalues of the Hamiltonian operator ${\hat H}$. It is best to work in the ``position" basis $X$. In this basis, the eigenfunctions $\psi_{E}(X)$ of $\hat H$ satisfies the standard Shr\"odinger equation,
\begin{equation}
-{1\over {2}}{ {d^2 \psi_{E}(X)}\over {dX^2}} + {\beta^2\over {2}} e^{\beta X} \psi_E(X)= E\psi_E(X),
\label{seq1}
\end{equation} 
valid in the range $-\infty < X<\infty$. It turns out that this Shr\"odinger equation has no bound state ($E<0$) and only has scattering states with $E\geq 0$. We label these positive energy eigenstates by $E= \beta^2k^2/8$, where $k$ is a continuous label varying from $0$ to $\infty$. A negative $k$ eigenfunction is the same as the positive $k$ eigenfunction, and hence it is counted only once. With this labeling, it turns out that the differential equation can be solved and one finds that the eigenfunction $\psi_k(X)$ is given by,
\begin{equation}
\psi_k(X)= a_k K_{ik}\left(2e^{\beta X/2}\right),
\label{sol1}
\end{equation}
where $K_{\nu}(y)$ is the modified Bessel function with index $\nu$. Note that, out of two possible solutions of the differential equation, we have chosen the one which does not diverge as $X\to \infty$, one of the physical boundary conditions. The important question is: how to determine the constant $a_k$ in Eq. (\ref{sol1})? Note that, unlike a bound state, the wavefunction $\psi_k(X)$ is not normalizable. To determine the constant $a_k$, we examine the asymptotic behavior of the wavefunction in the regime $X\to -\infty$. Using the asymptotic properties of the Bessel function (when its argument $2e^{\beta X/2}\to 0$), we find that\begin{equation}
\psi_k(X) \to a_k\left[ {{\Gamma(ik)}\over {2}}e^{-ik\beta X/2} - {{\pi}\over {2\sin(ik\pi)\Gamma(1+ik)}} 
e^{ik\beta X/2}\right].
\label{sol2}
\end{equation}
On the other hand, in the limit $X\to -\infty$, the Schr\"odinger equation (\ref{seq1}) reduces to a free problem,
\begin{equation}
-{1\over {2}}{ {d^2 \psi_{k}(X)}\over {dX^2}} = {{\beta^2 k^2}\over {8}}\psi_k(X),
\label{seq2}
\end{equation}
which allows plane wave solutions of the form,
\begin{equation}
\psi_k(X) \approx {\sqrt{\beta \over {4\pi}}}\left[e^{ik\beta X/2} + r(k) e^{-ik\beta X/2}\right],
\label{sol3}
\end{equation}
where $e^{ik\beta X/2}$ represents the incoming wave from $X=-\infty$ and $e^{-ik\beta X/2}$ represents the reflected wave going back towards $X=-\infty$ with $r(k)$ being the reflection coefficient. The amplitude ${\sqrt{\beta \over {4\pi}}}$ is chosen such that the plane waves $\psi_k(X)= \sqrt{\beta \over {4\pi}}e^{ik\beta X/2}$ are properly orthonormalized in the sense that $<\psi_k|\psi_k'>=\delta(k-k')$ where $\delta(z)$ is the Dirac delta function. Comparing Eqs. (\ref{sol2}) and (\ref{sol3}) in the regime $X\to -\infty$, we determine the constant $a_k$ (up to a phase factor),
\begin{equation}
a_k = \sqrt{ {\beta\over {\pi^3}}}\, {\sin(ik\pi)\Gamma(1+ik)}.
\label{ak1}
\end{equation}
The square of the amplitude $|a_k|^2$ (which is independent of the unknown phase factor) is then given by
\begin{equation}
|a_k|^2 = {{\beta k \sinh(\pi k)}\over {\pi^2}},
\label{ak2}
\end{equation}
where we have used the identity, $\Gamma(1+ik)\Gamma((1-ik)= \pi k/{\sinh (\pi k)}$. Therefore, the eigenstates of the operator $\hat H$ are given by $|k>$, such that ${\hat H}|k>={{\beta^2 k^2}\over 
{8}}|k>$ and in the $X$ basis, the wavefunction $\psi_k(X)=<k|X>$ is given (up to a phase factor) by the exact expression
\begin{equation}
\psi_k(X) = {{\sqrt{\beta k \sinh(\pi k)}}\over {\pi} } K_{ik}(2e^{\beta X/2}).
\label{eigen1}
\end{equation} 
We now go back to Eq. (\ref{pi3}) where we are ready to evaluate the matrix element $<X|e^{-{\hat H} L}|X>$ given the full knowledge of the eigenstates of $\hat H$. Expanding all the kets and bras in the eigenbasis $|k>$ of $\hat H$, we can rewrite Eq. (\ref{pi3}) as follows,
\begin{eqnarray}
{\overline {\rho^n(r)}}=A\, b_n \int_{-\infty}^{\infty} dX\, \int_0^{\infty} dk\, <X|k><k|X> \nonumber \\
e^{n\beta X}\, e^{-\beta^2 k^2 L/8} \nonumber \\
 = A\, b_n \int_0^{\infty} dk\, e^{-\beta^2 k^2 L/8} \int_{-\infty}^{\infty} dX |\psi_k(X)|^2 
e^{n\beta
X}.
\label{momn1}
\end{eqnarray}
The $X$ integral on the right hand side of Eq. (\ref{momn1}) can be expressed in a compact notation,
\begin{equation}
\int_{-\infty}^{\infty} dX |\psi_k(X)|^2 e^{n\beta
X}= <k|e^{n\beta {\hat X}}|k>.
\label{com1}
\end{equation}
Substituting the exact form of $\psi_k(X)$ from Eq. (\ref{eigen1}), we get
\begin{equation}
 <k|e^{n\beta {\hat X}}|k> = {{k\sinh(\pi k)}\over {\pi^2 2^{2n-1}}}\int_0^{\infty} dy y^{2n-1} 
K_{ik}(y)K_{-ik}(y).
\label{com2}
\end{equation}
Fortunately, the integral on the right hand side of Eq. (\ref{com2}) can be done in closed form ~\cite{grad1} and we obtain,
\begin{equation}
 <k|e^{n\beta {\hat X}}|k> = {{k\sinh(\pi k)}\over {\pi^2 2^{2n-1}}} {{\Gamma^2(n)}\over 
{\Gamma(2n)}}\Gamma(n-ik)\Gamma(n+ik).
\label{com3}
\end{equation}
Substituting this matrix element back in Eq. (\ref{momn1}), we arrive at our final expression,
\begin{eqnarray}
{\overline {\rho^n(r)}} & = &
A {{\beta^{2n+1}}\over {4\pi^2 2^{n}}} {{\Gamma(n)}\over {\Gamma(2n)}}\int_0^{\infty} dk\,k \sinh(\pi k) \nonumber \\
& & |\Gamma(n-ik)|^2  e^{-\beta^2 k^2 L/8}.
\label{fin1}
\end{eqnarray} 
To determine the constant $A$, we first put $n=1$ in Eq. (\ref{fin1}). Note that ${\overline {\rho(r)}}=1/L$ by virtue of the probability sum rule, $\int_0^L \rho(r)dr=1$. Taking the disorder average and using the translational invariance, one gets ${\overline {\rho(r)}}=1/L$. Using the identity, $\Gamma(1+ik)\Gamma((1-ik)= \pi k/{\sinh (\pi k)}$ and performing the integral on the right hand side
of Eq. (\ref{fin1}) and then demanding that the right hand side must equal $1/L$ for $n=1$, we get 
\begin{equation}
A = {\sqrt {2\pi L}}.
\label{a1}
\end{equation}
One can also check easily that $n\to 0$, the right hand side of Eq. (\ref{fin1}) approaches to $1$ as it should. In verifying this limit, we need to use the fact that $\Gamma(x)\approx 1/x$ as $x\to 0$ and also the identity, $\Gamma(ik)\Gamma(-ik)= \pi/{k \sinh (\pi k)}$. Now, for $n>0$ (strictly), one can make a further simplification of the right hand side of Eq. (\ref{fin1}). We use the property of the Gamma function, $\Gamma(x+1)=x\Gamma(x)$, repeatedly to write $\Gamma(n-ik)= (n-1-ik)\Gamma(n-1-ik)= (n-1-ik)(n-2-ik)\dots (1-ik)\Gamma(1-ik)$. Note that this formula, so far, is valid only for integer $n\geq 1$. This gives, for integer $n\geq 1$
\begin{eqnarray}
\Gamma(n-ik)\Gamma(n+ik) & = &
[(n-1)^2+k^2][(n-2)^2+k^2] ... \nonumber \\
 & & [1+k^2]\frac{\pi k}{\sinh(\pi k)}
\label{gamma1}
\end{eqnarray}
where we have used the identity, $\Gamma(1+ik)\Gamma((1-ik)= \pi k/{\sinh (\pi k)}$. Substituting this expression in Eq. (\ref{fin1}) we get, for $n\geq 1$,
\begin{eqnarray} 
{\overline {\rho^n(r)}} & = & 
\sqrt{2\pi L} {{\beta^{2n+1}}\over {4\pi 2^{n}}}{{\Gamma(n)}\over {\Gamma(2n)}}
\int_0^{\infty} dk\, k^2[(n-1)^2+k^2] \nonumber \\
 & & [(n-2)^2+k^2]\dots[1+k^2] e^{-\beta^2 k^2 L/8}.
\label{momn2}
\end{eqnarray}
Making the change of variable $\beta^2k^2 L/8 =z$ in the integral, we finally obtain the following expression for all integer $n\geq 1$,
\begin{eqnarray}
{\overline {\rho^n(r)}} & = &  
{1\over {L\sqrt {\pi}} }{ {\beta^{2n-2}} \over {2^{n-2}} }{ {\Gamma(n)}\over 
{\Gamma(2n)} }\int_0^{\infty} dz\, e^{-z} z^{1/2}\, \left[1^2+{{8z}\over {\beta^2 L}}\right] \nonumber \\
 & & \left[2^2+{{8z}\over {\beta^2 L}}\right]
\dots \left[(n-1)^2 + {{8z}\over {\beta^2 L}}\right].
\label{momn3}
\end{eqnarray}

For example, consider the case $n=2$. In this case, the formula in Eq.~(\ref{momn3}) gives
\begin{equation}
{\overline {\rho^2(r)}}= {{\beta^2}\over {12 L}}\left[1 +{ {12}\over {\beta^2 L}}\right],
\label{n2}
\end{equation}
which is valid for all $L$ and not just for large $L$. Note that the second term on the right hand side gives a contribution which is exactly $1/L^2$. This means that the variance, ${\overline {\rho^2(r)}}-{\overline {\rho(r)}}^2= \beta^2/{[12 L]}$ for all $L$. For arbitrary integer $n\geq 1$, taking the large $L$ limit in Eq.~(\ref{momn3}) we get, as $L\to \infty$,
\begin{equation}
{\overline {\rho^n(r)}} \to {1\over {L}} \left[ { {\beta^{2n-2}} \over {2^{n-2}} }{ {\Gamma^3(n)}\over
{\Gamma(2n)} }\right].
\label{genn}
\end{equation}.
Note that even though this expression was derived assuming integer $n\geq 1$, after obtaining this formula, one can analytically continue it for all noninteger $n>0$. Now, let us denote ${\rm Prob}(\rho,L)=P(\rho,L)$. Then ${\overline {\rho^n(r)}}=\int_0^{\infty} \rho^n P(\rho,L)d\rho$. Note again that the range of $\rho$ is from $0$ to $\infty$, since it is a probability density, and not a probability. The factor $1/L$ on the right hand side of Eq.~(\ref{genn}) suggests
that $P(\rho,L)$ has the following behavior for large $L$,
\begin{equation}
P(\rho,L) = {1\over {L}} f(\rho), 
\label{pyl}
\end{equation}
where the function $f(y)$ satisfies the equation,
\begin{equation}
\int_0^\infty y^n f(y) dy = \left[ { {\beta^{2n-2}} \over {2^{n-2}} }{ {\Gamma^3(n)}\over
{\Gamma(2n)} }\right].
\label{momn4}
\end{equation}
To determine $f(y)$ from this equation, we first use the identity, $\Gamma(2n)= 2^{2n-1}\Gamma(n)\Gamma(n+1/2)/{\sqrt{\pi}}$, known as the doubling formula for the Gamma function. Next we use ~\cite{grad2},
\begin{equation}
\int_0^{\infty} x^{n-1}e^{-ax}K_0(ax)dx =  {{\sqrt{\pi}}\over {(2a)^{n}}}{{\Gamma^2(n)}\over 
{\Gamma(n+1/2)}}.
\label{id1}
\end{equation}
Identifying the right hand side of Eq. (\ref{id1}) with the right hand side of Eq. (\ref{momn4})
upon choosing $a=2/{\beta^2}$, we get the exact expression of $f(y)$,
\begin{equation}
f(y) = {2\over {\beta^2 y}} e^{-2y/\beta^2}K_0\left({{2y}\over {\beta^2}}\right). 
\label{fy1}
\end{equation}
More cleanly, we can then write that for large $L$,
\begin{equation}
P(\rho, L) = {4\over {\beta^4 L}} f'\left[ {{2\rho}\over {\beta^2}}\right],
\label{scaled1}
\end{equation}
where the scaling function $f'(y)$ is universal (independent of the system parameter $\beta$) and is given 
by,
\begin{equation}
f'(y) = { {e^{-y}}\over {y} } K_0(y).
\label{gy1}
\end{equation}
This function has the following asymptotic behaviors,
\begin{equation}
f'(y) \approx \cases 
             { {1\over {y}}\left[-\ln(y/2)+0.5772\ldots\right], \,\,\,      &$y\to 0$, \cr
               \sqrt{ {{\pi}\over {2y^3}}} e^{-2y}, \,\,\,  &$y\to \infty$. \cr}
\label{gy2}
\end{equation}

The scaling form in Eq.~(\ref{scaled1}) is valid only when $m(r)\sim L$. If $m(r)$ is a number of order $O(1)$ (not as large as $L$), then the scaling breaks down. This fact suggests that the correct behavior of the distribution $P(\rho, L)$ for large $L$ actually has two parts,

\begin{equation}
P(\rho,L) \approx \left[1- {{\ln^2 (L)}\over {\beta^2 L}}\right]\delta(\rho) + 
{4\over {\beta^4 L}} f'\left[ {{2\rho}\over {\beta^2}}\right]\theta\left(\rho-{c\over {L}}\right),
\label{scaled2}
\end{equation}

where $f'(y)$ is given by Eq.~(\ref{gy1}). This form in Eq.~(\ref{scaled2}) is consistent with all the observed facts. For example, if one integrates the right hand side, the first term gives $1- {{\ln^2 (L)}\over {\beta^2 L}}$ (with the convention $\int_0^{\infty}\delta(y)dy=1$). The second term, when integrated, gives ${{\ln^2 (L)}\over {\beta^2 L}}$ (where we have used the small $y$ behavior of $f'(y)$ from Eq.~(\ref{gy2}) and kept only the leading order term for large $L$) which exactly cancels the identical factor in the first term to give a total sum $1$, as it should. On the other hand, for any finite moment of order $n$, the first term does not contribute and only the second term contributes to give the result in Eq.~(\ref{genn}). 

\subsection{ The Density-Density Correlation Function} 
We now consider the density-density correlation function between two points $r_1$ and $r_2$ at equilibrium. The calculation proceeds more or less along the same lines as in the previous section.
The density-density correlation function is defined as 
\begin{equation}
C(r_1,r_2) = {\overline {\rho(r_1)\rho(r_2)}},
\label{corr1}
\end{equation}
which evidently depends only on $r=|r_1-r_2|$ due to the translational invariance. The density $\rho(r)$ is again given by Eq.~(\ref{Gibbs2}). It follows from Eq.~(\ref{Gibbs2}) that
\begin{equation}
\rho(r_1)\rho(r_2) = {{e^{\beta\left[h(r_1)+h(r_2)\right]}}\over {Z^2}}=\int_0^{\infty} dy\, 
y e^{-yZ + \beta[h(r_1)+h(r_2)]},
\label{corr2}
\end{equation}
where the partition function, $Z=\int_0^L dr e^{\beta U(r)}$ and we have used the identity, $1/{Z^2}=\int_0^{\infty} dy\,y e^{-Zy} $. As in Section-II, we now make a change of variable in Eq.~(\ref{corr2}) by writing $y= \beta^2 e^{\beta u}/2$. Then Eq. (\ref{corr2}) becomes,
\begin{eqnarray}
\rho(r_1)\rho(r_2) & = &
{{\beta^5}\over {4}}\int_{-\infty}^{\infty} du\, \exp [
-{{\beta^2}\over {2}}\left\{ \int_0^{L} dx e^{\beta [h(x)+u]}\right\} + \nonumber  \\
 & & \beta (h(r_1)+u + h(r_2)+u)],
\label{corr3}
\end{eqnarray}
where we have used the explicit expression of the partition function, $Z=\int_0^{L} dr e^{\beta h(r)}$. Averaging over the disorder, we get
\begin{eqnarray}
{\overline {\rho(r_1)\rho(r_2)}} & = &
B {{\beta^5}\over {4}}\int_{-\infty}^{\infty} du\, \int_{h(0)=0}^{h(L)=0} {\cal D} h(x)\, \exp[-\left\{\int_0^{L} dx\, \right.  \nonumber \\
 & & \left. \left[{1\over {2}}{\left( {{dh(x)}\over
{dx}}\right)}^2[{1\over {2}}{\left( {{dh(x)}\over
{dx}}\right)}^2 +{{\beta^2}\over {2}}e^{\beta(h(x)+u)}\right]\right\}  \nonumber \\
 & & +\beta(h(r_1)+h(r_2)+2u)]       
\label{avcorr1}
\end{eqnarray}
where the normalization constant $B$ will be determined from the condition,
$\int_0^{L}\int_0^{L} C(r_1,r_2)dr_1dr_2 =1$ (which follows from the
fact that $\int_0^L \rho(r)dr =1$). Alternatively, one can put $r=r_2-r_1=0$
in the expression for the correlation function and then it should be same
as ${\overline {\rho^2(r)}}$ already computed in the previous section.\\

As before, we next shift the potential, i.e., we define $V(x)=U(x)+u$ for all $x$. The Eq.~(\ref{avcorr1}) then simplifies,
\begin{eqnarray}
{\overline {\rho(r_1)\rho(r_2)}} & = &
B {{\beta^5}\over {4}}\int_{-\infty}^{\infty} du\, \int_{V(0)=u}^{V(L)=u} {\cal D} V(x)\, \nonumber \\
 & & \exp\left[-\left\{\int_0^{L} dx\, \left[{1\over {2}}{\left( {{dV(x)}\over
{dx}}\right)}^2 + \nonumber \right. \right. \right. \\
 & & \left. \left. \left. {{\beta^2}\over {2}}e^{\beta V(x)}\right]\right\}
+ \beta(V(r_1) + V(r_2))\right].
\label{avcorr2}
\end{eqnarray}

Thus we have again reduced the problem to a path integral problem. However, there is a difference in the subsequent calculations. This is because, unlike the previous calculation, we now have to divide the paths into $3$ parts: (i) paths
starting at $V(0)=u$ and propagating up to the point $r_1$ where $V(r_1)=V_1$ (note that $V_1$ can vary from $-\infty$ to $\infty$), (ii) paths starting at $r_1$ with $V(r_1)=V_1$ and propagating up to $r_2$ with $V(r_2)=V_2$ and (iii) paths starting at $r_2$ with $V(r_2)=V_2$ and propagating up to $L$ where $V(L)=u$. We have assumed $r_2\geq r_1$ for convenience. Using the bra-ket notation, we can then re-write Eq. (\ref{avcorr2}) as
\begin{eqnarray}
{\overline {\rho(r_1)\rho(r_2)}} & = & 
B {{\beta^5}\over {4}}\int_{-\infty}^{\infty} du\,\int_{-\infty}^{\infty} dV_1\, \int_{-\infty}^{\infty} dV_2 \, \nonumber \\
 & & <u|e^{-{\hat H}r_1}|V_1>e^{\beta V_1}
<V_1|e^{-{\hat H}(r_2-r_1)}|V_2> \nonumber \\
 & & e^{\beta V_2} <V_2|e^{-{\hat H}(L-r_2)}|u>.
\label{avcorr3}
\end{eqnarray}
The Hamiltonian ${\hat H}\equiv {1\over {2}}\left({{dV}\over {dx}}\right)^2 + {{\beta^2}\over {2}}e^{\beta V(x)}$ is the same as in the previous section. Using $\int_{-\infty}^{\infty} du\, |u><u| = {\hat I}$, Eq. (\ref{avcorr3}) can be simplified,
\begin{eqnarray}
{\overline {\rho(r_1)\rho(r_2)}} & = &
B {{\beta^5}\over {4}} \int_{-\infty}^{\infty} dV_1\, \int_{-\infty}^{\infty} dV_2 \, <V_2|e^{-{\hat H}(L-r)}|V_1> \nonumber \\
 & & <V_1|e^{-{\hat H}r}|V_2> e^{\beta (V_1+V_2)},
\label{avcorr4}
\end{eqnarray}
where $r=r_2-r_1$. Note that Eq. (\ref{avcorr4}) clearly shows that $C(r_1,r_2,L)=C(r=r_2-r_1,L)$, as it should due to the translational invariance. Furthermore, Eq. (\ref{avcorr4}) also shows that that function $C(r,L)$ is symmetric around $r=L/2$, i.e., $C(r,L)=C(L-r,L)$. This last fact is expected due to the periodic boundary condition. As before, we change to a more friendly notation: $V_1\equiv X_1$ and $V_2\equiv X_2$, where $X_1$ and $X_2$ denote the `positions' of the fictitious quantum particle at `times' $r_1$ and $r_2$. With this notation, Eq. (\ref{avcorr4}) reads,
\begin{eqnarray}
{\overline {\rho(r_1)\rho(r_2)}} & = & B {{\beta^5}\over {4}}
\int_{-\infty}^{\infty} dX_1\, \int_{-\infty}^{\infty} dX_2 \,\nonumber \\
 & &<X_2|e^{-{\hat H}(L-r)}|X_1> \nonumber \\
 & & <X_1|e^{-{\hat H}r}|X_2> e^{\beta (X_1+X_2)}.
\label{avcorr5}
\end{eqnarray}

This can be solved to obtain the correlation function - 

\begin{eqnarray}
C(r,L) & = & B { {\beta^5}\over {256}}\int_0^{\infty}\int_0^{\infty} dk_1dk_2 k_1k_2 (k_1^2-k_2^2)^2 \nonumber \\ 
 & & { {\sinh(\pi k_1)\sinh(\pi k_2)}\over {[\cosh(\pi k_1)-\cosh(\pi k_2)]^2}} \nonumber \\
 & & \exp\left[-{{\beta^2}\over {8}} \left( k_1^2(L-r)+k_2^2 r \right) \right].
\label{corrfin1}
\end{eqnarray}
For $r=0$, it is possible to perform the double integral in Eq. (\ref{corrfin1})and one finds that it reduces to the expression of ${\overline {\rho^2(r)}}$ in Eq. (\ref{n2}) of the previous section, provided the normalization constant $B=\sqrt{2\pi L}$. Thus, the two-point density-density correlator is given exactly by Eq. (\ref{corrfin1}) (with $B=\sqrt{2\pi L}$) and note that this expression is valid for all $L$. This exact expression of the correlation function was first derived by Comtet and Texier~\cite{comtet} in the context of a localization problem in disordered supersymmetric quantum mechanics.\\ 

To extract the asymptotic behavior for large $L$, we rescale $k_1 \sqrt{L-r}=x_1$ and $k_2\sqrt{L}=x_2$ in Eq. (\ref{corrfin1}), then expand the $\sinh$'s and the $\cosh$'s for small arguments, perform the resulting double integral (which becomes simple after the expansion) and finally get for $L\to \infty$ and $r\neq 0$,
\begin{equation}
C(r,L) \to {1\over { \sqrt{2\pi \beta^2} L^{5/2} [x(1-x)]^{3/2}}},
\label{scorr1}
\end{equation}

where $x=r/L$ is the scaling variable.If we identify $\rho$ with $m/L$, we can identify the expressions for $P(\rho)$ and $C(r,L)$ with the corresponding equilibrium quantities - $P(n,L)=\frac{1}{L}P(\rho)$ and $G(r,L)=L^2C(r,L)$. So, for $n \geq 1$ and $r \geq 1$

\begin{equation}
P(n, L) = {4\over {\beta^4 L^2}} f'\left[ {{2 n}\over {\beta^2 L}}\right],
\label{connection1}
\end{equation}

and 

\begin{equation}
G(r,L) \to {1\over { \sqrt{2\pi \beta^2} L^{1/2} [x(1-x)]^{3/2}}},
\label{connection2}
\end{equation}

 We see that the scaling forms in these cases are similar. A fit to the functional forms shows that  these equilibrium results reproduce quite well the scaling exponents and scaling functions for $G(r)$ and $P(n)$ for $n \geq 1$ obtained numerically for the nonequilibrium case $\omega = K = 1$, as can be seen in Figs.~\ref{advcorr} and ~\ref{advdensity}, though with different values of $\beta$. The correlation function matches with $\beta \simeq 4$ while $\beta \simeq 2.3$ describes the probability distribution of number data well. However, $P(0,L)$ (and thus $N_{occ}$) does not agree closely in the two cases. The equilibrium case can also be used to shed light on the dynamical properties of the nonequilibrium steady state. We compared our results for $G(t,L)$ with the density-density autocorrelation function in the adiabatic $\omega \rightarrow 0$ limit. To find the latter, we simulated a surface with height field $h(r,t)$ evolving according to KPZ dynamics, and evaluated the density using the equilibrium weight $\rho(r,t)= e^{-\beta h(r,t)}/Z$. As shown in ~\cite{nagar}, the results with $\beta = 4$ agree with the autocorrelation function in the nonequilibrium system, apart from a numerical factor.\\

It is surprising that results in this equilibrium limit describes the non-equilibrium state so well. In the non-equilibrium case, the driving force behind particle motion  and clustering is the surface fluctuation fluctuation while the equilibrium case, it is the temperature. The common feature in both the cases is the surface terrain. Thus, in some region of parameter space the surface motion mimics temperature and causes the particles to redistribute in a certain way. Why the equivalent temperature for various quantities is different is not clear and deserves further study.\\

\section{v. FUTURE WORK}

In this paper, we have described our results on the problem of particles sliding on a KPZ surface, with both the surface and the particles moving in the same direction, corresponding to the case of particle advection by a noisy Burgers flow. We see that in the steady state, the two-point density-density correlation function diverges near the origin as a function of distance scaled with the system size. This is an indicator of strong clustering of particles and the defining characteristic of a new kind of state - the strong clustering state (SCS).\\

Questions arise about the robustness of the strong clustering state -  Does clustering survive in the case of anti-advection where the surface and particles move in opposite directions to each other? What happens if we change the symmetry properties of the driving field, and have driving by an Edwards-Wilkinson (EW) surface instead of the KPZ surface? Does the phenomenon survive in higher dimensions? These questions will be addressed in a subsequent paper~\cite{future}, where it will be shown that the steady state is of the SCS kind in all these cases, even though the degree of clustering differs from one case to another.

\section{ACKNOWLEDGEMENTS}

We thank A.Comtet for very useful discusssions. SNM and MB acknowledge support from the Indo-French Centre for the Promotion of Advanced Research (IFCPAR). AN acknowledges support from the Kanwal Rekhi Career Development Awards.


\begin{thebibliography}{40}
\bibitem[{\citenamefont{Drossel and Kardar}(2000)}]{drosselcomment}
 This is in contrast to Drossel and Kardar's update rule~\cite{drossel2} where only particles at a site affected by the surface evolution are moved.
\end{thebibliography}
\end{document}